\documentclass[runningheads]{llncs}
\usepackage{graphicx}
\usepackage{mathrsfs}
\usepackage{amsmath}
\usepackage{cite}
\usepackage{subcaption,graphicx}
\usepackage{float}
\usepackage[misc]{ifsym}
\usepackage{bbding}
\usepackage{dblfloatfix}
\usepackage[nodisplayskipstretch]{setspace}
\usepackage{siunitx}

\begin{document}
\title{CNN-based Cardiac Motion Extraction to Generate Deformable Geometric Left Ventricle Myocardial Models from Cine MRI}
\titlerunning{CNN-based Framework to Generate Geometric LV Myocardial Models}
\author{Roshan Reddy Upendra\inst{1}$^{\textrm{\Letter}}$ \and Brian Jamison Wentz\inst{3,5} \and
Richard Simon\inst{2}  \and Suzanne M. Shontz\inst{3,4,5}
Cristian A. Linte\inst{1,2}}
\authorrunning{R.R.Upendra et al.}
\institute{$^1$ Center for Imaging Science, 
 $^2$ Biomedical Engineering, Rochester Institute of Technology, Rochester, NY USA\\
  \email{ru6928@rit.edu}\\
 $^3$ Bioengineering Graduate Program,
$^4$ Electrical Engineering and Computer Science,
$^5$ Information and Telecommunication Technology Center, University of Kansas, Lawrence, KS, USA\\
}
\maketitle              % typeset the header of the contribution

\begin{abstract}

Patient-specific left ventricle (LV) myocardial models have the potential to be used in a variety of clinical scenarios for improved diagnosis and treatment plans.  Cine cardiac magnetic resonance (MR) imaging provides high resolution images to reconstruct patient-specific geometric models of the LV myocardium. With the advent of deep learning, accurate segmentation of cardiac chambers from cine cardiac MR images and unsupervised learning for image registration for cardiac motion estimation on a large number of image datasets is attainable. Here, we propose a deep leaning-based framework for the development of patient-specific geometric models of LV myocardium from cine cardiac MR images, using the Automated Cardiac Diagnosis Challenge (ACDC) dataset. We use the deformation field estimated from the VoxelMorph-based convolutional neural network (CNN) to propagate the isosurface mesh and volume mesh of the end-diastole (ED) frame to the subsequent frames of the cardiac cycle. We assess the CNN-based propagated models against segmented models at each cardiac phase, as well as models propagated using another traditional nonrigid image registration technique. 

\end{abstract}
\vspace{-0.75cm}
\section{Introduction}

To reduce the morbidity and mortality associated with cardiovascular diseases (CVDs) \cite{benjamin2017heart}, and to improve their treatment, it is crucial to detect and predict the progression of the diseases at an early stage. In a clinical set-up, population-based metrics, including measurements of cardiac wall motion, ventricular volumes, cardiac chamber flow patterns, etc., derived from cardiac imaging are used for diagnosis, prognosis and therapy planning. 

In recent years, image-based computational models have been increasingly used to study ventricular mechanics associated with various cardiac conditions. A comprehensive review of patient-specific cardiovascular modeling and its applications is described in \cite{smith2011euheart}. Cardiovascular patient-specific modeling includes a geometric representation of some or all cardiac chambers of the patient's anatomy and is derived from different imaging modalities \cite{gray2018patient}. 

The construction of patient-specific geometric models entails several steps: clinical imaging, segmentation and geometry reconstruction, and spatial discretization (i.e., mesh generation) \cite{morris2016computational}. For example, Bello \textit{et al.} \cite{bello2019deep} presented a deep learning based framework for human survival prediction for patients diagnosed with pulmonary hypertension using cine cardiac MR images. Here, the authors employ a 4D spatio-temporal B-spline image registration method to estimate the deformation field at each voxel and at each timeframe. The estimated deformation field was used to propagate the ED surface mesh of the right ventricle (RV), reconstructed from the segmentation map, to the rest of the timeframes of a particular subject. Cardiac MRI is a current gold standard to assess global (ventricle volume and ejection fraction) and regional (kinematics and contractility) function of the heart under various diseases. In particular, cardiac MRI enables the generation of high quality myocardial models, which can, in turn, be used to identify reduced function.

\begin{figure}[t]
\begin{center}
   \includegraphics[width=0.9\linewidth]{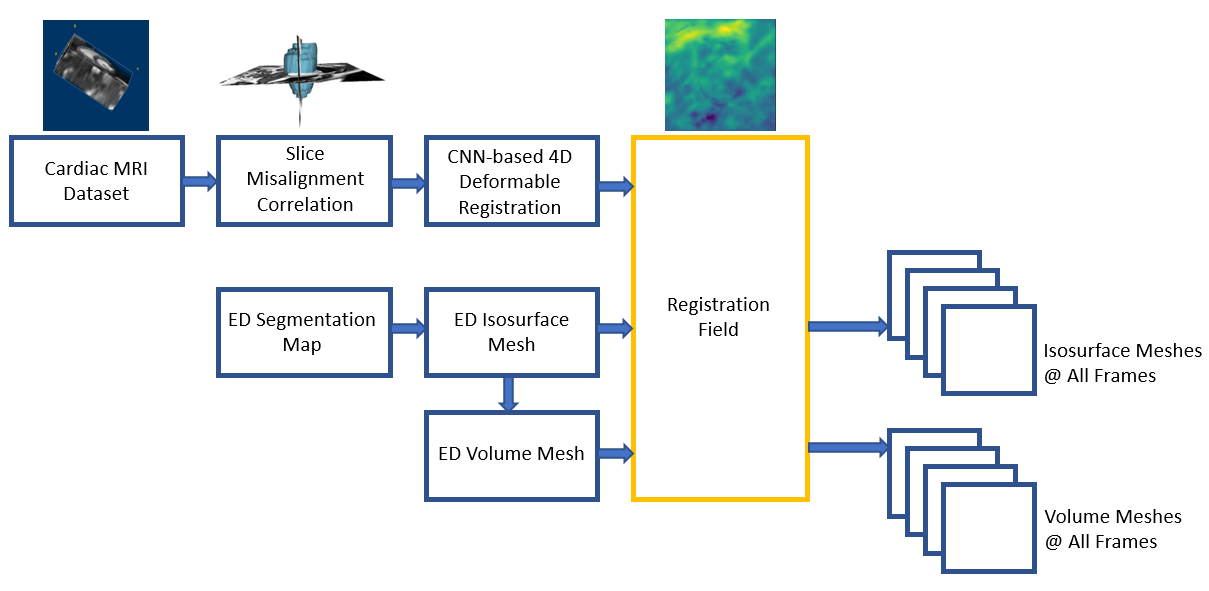}
\end{center}
   \caption{Overview of the proposed CNN-based workflow to generate patient-specific LV myocardial geometric model.}
\label{fig:m1}
\end{figure}

In this work, we propose a deep learning-based pipeline to develop patient-specific geometric models of the LV myocardium from cine cardiac MR images (Fig. \ref{fig:m1}). These models may be used to conduct various simulations, such as assessing myocardial viability. In our previous work \cite{upendra2020convolutional}, we introduced a preliminary, proof of concept, CNN-based 4D deformable registration method for cardiac motion estimation from cine cardiac MR images, using the ACDC dataset \cite{bernard2018deep}. Here, we demonstrate the use of the CNN-based 4D deformable registration technique to build dynamic patient-specific LV myocardial models across subjects with different pathologies, namely normal, dilated cardiomyopathy (DCM), hypertrophic cardiomyopathy (HCM) and subjects with prior myocardial infarctions (MINF). Following segmentation of the ED cardiac frame, we generate both isosurface and volume LV meshes, which we then propagate through the cardiac cycle using the CNN-based registration fields. In addition, we demonstrate the generation of dynamic LV volume meshes depicting the heart at various cardiac phases by warping a patient-specific ED volume mesh based on the registration-based propagated surface meshes. Lastly, we compare these meshes to those obtained by directly propagating the ED volume mesh using the CNN-based deformation fields.

\vspace{-0.25cm}
\section{Methodology}

\vspace{-0.15cm}
\subsection{Cardiac MRI Data}

We use the 2017 ACDC dataset that was acquired from real clinical exams. The dataset is composed of cine cardiac MR images from $150$ subjects, divided into five equally-distributed subgroups: normal, MINF, DCM, HCM and abnormal RV. The MR image acquisitions were obtained using two different MR scanners of $1.5$ T and $3.0$ T magnetic strength. These series of short axis slices cover the LV from base to apex such that one image is captured every $5$ {mm} to $10$ {mm} with a spatial resolution of $1.37$ mm\textsuperscript{2}/pixel to $1.68$ mm\textsuperscript{2}/pixel. 

\vspace{-0.3cm}
\subsection{Image Preprocessing}
\label{section:sma}

We first correct for the inherent slice misalignments that occur during the cine cardiac MR image acquisition. We train a modified version of the U-Net model \cite{ronneberger2015u} to segment the cardiac chambers, namely LV blood-pool, LV myocardium and RV blood-pool, from 2D cardiac MR images. We identify the LV blood-pool center, i.e., the centroid of the predicted segmentation mask and stack the 2D cardiac MR slices collinearly to obtain slice misalignment corrected 3D images \cite{upendra2020convolutional, dangi2018cine}. 

\vspace{-0.3cm}
\subsection{Deformable Image Registration}

\subsubsection{CNN-based Image Registration.} 

We leverage our 4D deformable registration method described in \cite{upendra2020convolutional} which employs the VoxelMorph  \cite{balakrishnan2019voxelmorph} framework to determine the optical flow representation between the slice misalignment corrected 3D images. The CNN is trained using the following loss function: 
\begin{equation}
{L} = {L}_\text{similarity} + \lambda{L}_\text{smooth}, \label{eqn:1}
\end{equation}
where ${L}_{\text{similarity}}$ is the mean squared error (MSE) between the target frame and the warped frame, ${L}_{\text{smooth}}$ is the smoothing loss function to spatially smooth the registration field, and $\lambda$ is the regularization parameter, which is set to $10^{-3}$ in our experiments. Inspired by Zhu \textit{et al.} \cite{zhu2020new}, we use the Laplacian operator in the smoothing loss function as it considers both global and local properties of the objective function $y = x^2$ instead of the traditional gradient operator which considers only the local properties of the function $y = x^2$. A detailed comparison of both these smoothing loss functions with respect to cardiac motion estimation from cine MR images is found in \cite{upendra2020convolutional}.

The 4D cine cardiac MRI datasets are composed of $28$ to $40$ 3D image frames that cover the complete cardiac cycle. For this discussion, we shall refer to the 3D images as $I_{ED}$, $I_{ED+1}$,...,$I_{ED+N_{T}-1}$ where $I_{ED}$ is the end-diastole image frame, and $N_T$ is the total number of 3D images. We employ the fixed reference frame registration method, wherein the task is to find an optical flow representation between the image pairs $\{(I_{ED}, I_{ED+t})\}_{t=1,2,3,...,N_{T}-1}$.

During training, we use $110$ of the total $150$ MR image dataset for training, $10$ for validation and the remaining $30$ for testing. The CNN for cardiac motion estimation is trained using an Adam optimizer with a learning rate of $10^{-4}$, halved at every $10^{th}$ epoch for $50$ epochs. Both, the U-Net model used for slice misalignment correction and VoxelMorph network trained to estimate cardiac motion were trained on  NVIDIA RTX 2080 Ti GPU. 
 
\vspace{-0.15cm}
 
\subsubsection{Conventional Image Registration.} We compare the performance of the VoxelMorph framework with that of the B-spline free form deformation (FFD) nonrigid image registration algorithm  \cite{rueckert1999nonrigid}. This iterative intensity-based image registration method was implemented using SimpleElastix \cite{marstal2016simpleelastix, klein2009elastix}, which enables a variety of image-registration algorithms in different programming languages. The FFD algorithm was set to use the adaptive stochastic gradient descent method as the optimizer, MSE as the similarity measure and binding energy as the regularization function. The FFD-based image registration was optimized in $500$ iterations, while sampling $2048$ random points per iteration, on an Intel(R) Core(TM) i9-9900K CPU. 

\subsection{Mesh Generation and Propagation}
\label{section: mesh}

We use the manual segmentation map of the ED frame to generate isosurface meshes. The slice thickness of each MRI image slice is $5$ {mm} to $10$ {mm}, however, in order to obtain good quality meshes, the segmentation maps were resampled to a slice thickness of $1$ {mm}. We use the Lewiner marching cubes \cite{lewiner2003efficient} algorithm to generate the meshes from the resampled segmentation maps of the ED frames, and then simplification techniques, such as vertex simplification and edge collapse, were performed using MeshLab $2020.07$ \cite{cignoni2008meshlab}. The simplification techniques are repeated multiple times to reduce the number of vertices until the mesh has been fully decimated while preserving the anatomical integrity and aspect ratio of the isosurface meshes. 

Volume meshes of the initial surface meshes at the end-diastolic phases for four patients with various heart conditions were generated based on the decimated patient-specific surface meshes using Tetgen 1.6 \cite{si2015tetgen}. In particular, a constrained Delaunay mesh generation algorithm was used to generate tetrahedral meshes based on the triangulated surface meshes. Steiner points were added within the boundary of the surface mesh so that the tetrahedra maintained a radius-edge ratio of $1.01$ and a maximum volume of $9$ mm\textsuperscript{3} as needed for generation of valid meshes \cite{si2015tetgen}.

Mesh quality assessment was performed on the ED volume meshes utilizing the scaled Jacobian metric, which ranges between $-1$ to $+1$, where $+1$ indicates an ideal equilateral tetrahedron, while negative and zero scaled Jacobian values indicate inverted and degenerate tetrahedral elements, respectively. Tetrahedra with a scaled Jacobian greater than or equal to $0.2$ are considered acceptable \cite{knupp2003algebraic}. The ED volume mesh has a minimum scaled Jacobian value of $0.078$, which demonstrates a valid, non-tangled mesh. However, the end-systole (ES) phase mesh contains some lower quality elements indicated by lower minimum scaled Jacobian values.

\begin{figure}
\begin{center}
   \includegraphics[width=1.0\linewidth]{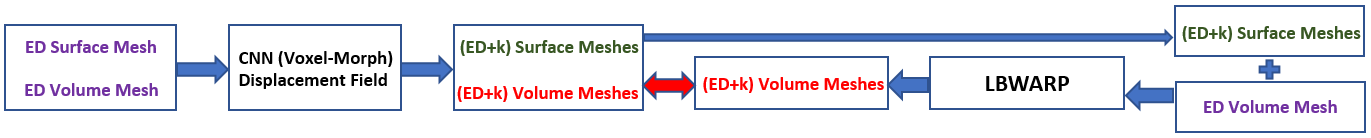}
\end{center}
   \caption{Pipeline to generate dynamic volume meshes (at cardiac frames (ED + k)) by direct CNN-based propagation, as well as volume mesh warping based on dynamic boundary meshes.}
\label{fig:m2}
\end{figure}

To demonstrate the VoxelMorph-based motion extraction and propagation to build patient-specific LV myocardial models, we generate two sets of volume meshes at each cardiac frame for each patient in each pathology group (Fig. \ref{fig:m2}).

\vspace{-0.3cm}

\begin{figure}
\begin{center}
   \includegraphics[width=0.7\linewidth]{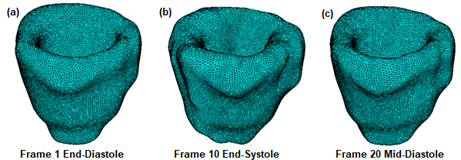}
\end{center}
   \caption{LV volume meshes at three cardiac phases (a) end-diastole; (b) end-systole; and (c) mid-diastole generated using LBWARP.}
\label{fig:brian}
\end{figure}

\vspace{-0.4cm}
The first set is produced by propagating the volume meshes at the ED frame to all the subsequent frames of the cardiac cycle using the deformation field estimated by the VoxelMorph-based registration method. For the second set, the ED volume mesh generated with Tetgen was used to generate the volume meshes corresponding to the other cardiac phases. We employed the log barrier-based mesh warping (LBWARP) method \cite{shontz2003mesh} to deform the ED volume mesh onto the target surface mesh for the new cardiac phase (Fig. \ref{fig:brian}). The method computes new positions for the interior vertices in the ED volume mesh, while maintaining the mesh topology and point-to-point correspondence \cite{shontz2003mesh}.

\vspace{-0.3cm}

\vspace{-0.4cm}

Briefly, LBWARP first calculates a set of local weights for each interior vertex in the initial (ED) volume mesh based on the relative inverse distances from each of its neighbors, which specify the representation of each interior vertex in terms of its neighbors. Next, the vertices in the ED surface mesh are mapped onto the new surface boundary. Finally, the interior vertices in the ED volume mesh are then repositioned to reflect the updated positions of the boundary nodes, while maintaining edge connectivity and point-to-point correspondence, and ultimately yielding the volume meshes that correspond to each new cardiac phase.

\vspace{-0.3cm}
\section{Results and Discussion}
\vspace{-0.2cm}

To evaluate the registration performance, the LV isosurface (generated from the ED image segmentation map) is propagated to all the subsequent cardiac frames using the deformation field estimated by FFD and VoxelMorph. We then compare these isosurfaces to those directly generated by segmenting all cardiac image frames using a modified U-Net model \cite{ronneberger2015u} (Section \ref{section:sma}), which we refer to as the ``silver standard".

Table \ref{table: dice} summarizes the performance of the FFD and VoxelMorph registration by assessing the Dice score and mean absolute distance (MAD) between the propagated and directly segmented (i.e., ``silver standard'') isosurfaces. 

\begin{figure*}[h!]
      \includegraphics[width=11.0cm]{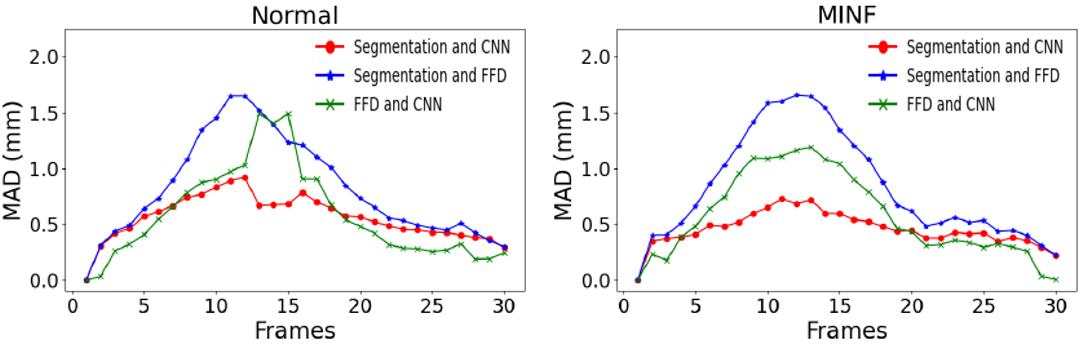}
      \includegraphics[width=11.1cm]{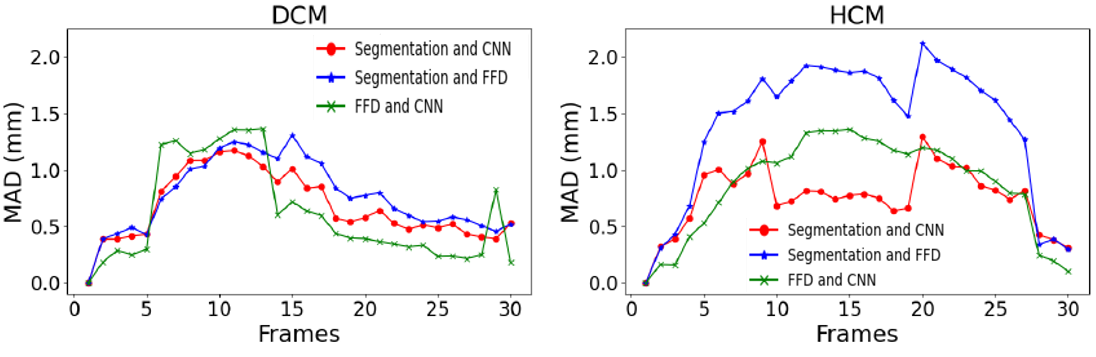}
  \caption {MAD between FFD- and CNN-propagated, and segmented (i.e., ``silver standard'') isosurfaces at all cardiac frames for all patient pathologies.}
\label{fig:diceMadPlots}
\end{figure*}

Fig. \ref{fig:diceMadPlots} illustrates the distance between the three sets of isosurfces (segmented, CNN-propagated and FFD-propagated) for one patient from each pathology. The MAD between the surfaces is less than 2 mm at all frames, with the CNN-propagated isosurfaces being closest to the ``silver standard'' segmented surfaces.
% \vspace{-0.3cm}

\begin{table*}[b]

\caption{Mean Dice score (\%) and mean absolute distance (MAD) (mm) between FFD and segmentation (FFD-SEG), CNN and segmentation (CNN-SEG), and FFD and CNN (FFD-CNN) results. Statistically significant differences were evaluated using the t-test (* for p $<$ 0.1 and ** for p $<$ 0.05).}
\begin{center}
\begin{tabular}{|c|l|l|l|l|l|l|l|l|r|}
\hline
  &  \multicolumn{2}{|c|}{Normal} & \multicolumn{2}{|c|}{MINF} & \multicolumn{2}{|c|}{DCM} & \multicolumn{2}{|c|}{HCM} \\
\hline
   & Dice & MAD & Dice & MAD & Dice & MAD & Dice & MAD \\
\hline
  FFD-Segmentation & 74.80 & 1.53 & 77.69 & 1.09 & 80.41 & 0.91 & 77.39 & 1.97\\
\hline
 CNN-Segmentation & {80.41**} & 1.15 & {81.21*} & 0.87 & {83.39*} & 0.91 & {82.46*} & 1.09\\
\hline
 FFD-CNN & 77.81 & 1.13 & 82.12 & 0.75 & 81.67 & 0.97 & 77.34 & 1.77\\
\hline
\end{tabular}
\end{center}
\label{table: dice}
\end{table*}

% \vspace{-1.5cm}

\begin{figure*}[h!]
      \includegraphics[width=5.8cm]{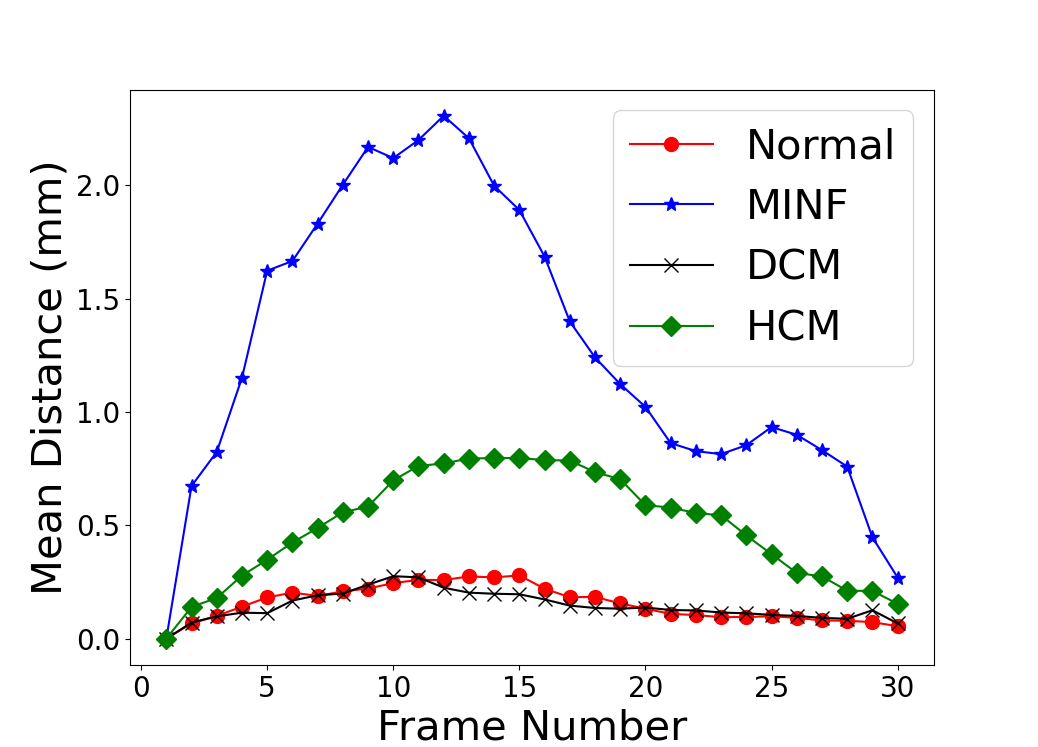}
      \includegraphics[width=5.8cm]{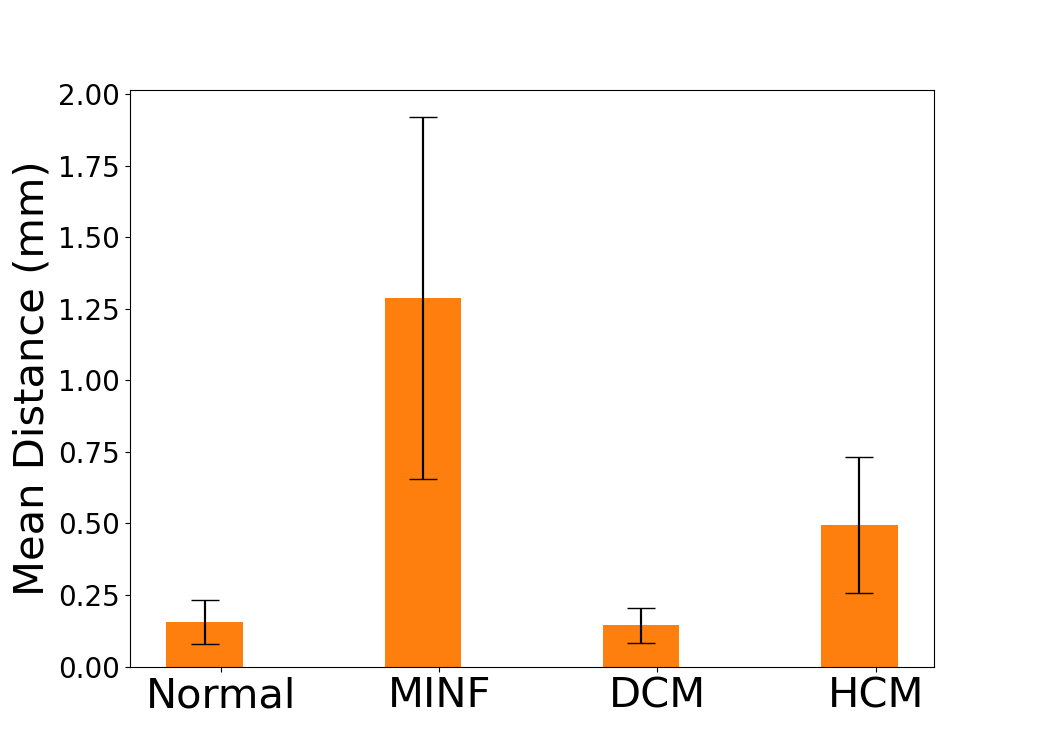}
  \caption {Mean node-to-node distance at each cardiac frame between the CNN-propagated and LBWARP-generated volume meshes (left); mean (std-dev) node distance across all frames for each patient pathology (right).}
\label{fig:volPlots}
\end{figure*}

\vspace{-0cm}
As mentioned in Section \ref{section: mesh} and shown in Fig. \ref{fig:m2}, we generate two sets of volume meshes at each frame of the cardiac cycle. Fig. \ref{fig:volPlots} shows the mean node distance between the two sets of volume meshes across all cardiac frames for one subject in each of the four pathologies. Fig. \ref{fig:volPlots} shows the mean node distance between the two sets of volume meshes at each frame of the cardiac cycle for the four subjects. It can be observed that the two sets of volume meshes are in close agreement with each other, exhibiting a mesh-to-mesh distance within 2 mm.

% \begin{table*}[tb]

% \caption{Mean (std-dev) node distance between the isosurface meshes and volume meshes for all cardiac phases compared between FFD and VoxelMorph-based propagation techniques.}
% \begin{center}
% \begin{tabular}{|c|c|l|l|l|l|l|l|l|r|}
% \hline
%  & Normal & MINF & DCM & HCM  \\
% \hline
% ED + $5^{th}$ Frame & 0.21 & 1.66 & 0.16 & 0.42  \\
% \hline
% ES Frame & 0.25 & 2.31 & 0.22 & 0.79 \\
% \hline
% ES + $5^{th}$ Frame & 0.18 & 1.39 & 0.14 & 0.71 \\
% \hline
% ES + $10^{th}$ Frame & 0.11 & 0.82 & 0.12 & 0.45 \\
% \hline
% \end{tabular}
% \end{center}
% \label{table: meshFfdVm}
% \end{table*}

% One of the major advantage of the proposed CNN-based pipeline to develop patient-specific geometric model is the computational time required. For example, it takes around $90$ seconds to propagate a volume mesh at ED frame to the other frames of the cardiac cycle using a trained VoxelMorph model, compared to $235$ seconds using the FFD-based registration method. 

We also briefly investigated the effect of using initial-to-final frame vs. adjacent frame-to-frame registration to extract the cardiac motion throughout the cycle. Although the sequential registration method estimates smaller deformation between two consecutive, adjacent image frames compared to the larger deformations estimated by the initial-to-final frame registration, their concatenation across several frames accumulates considerable registration errors. As such, when using these concatenated registration-predicted deformation fields to propagate the ED isosurfaces and volume meshes to the subsequent cardiac phases, the Dice score and MAD between the propagated and segmented geometries rapidly deteriorate, along with the quality of the propagated surface and volume meshes.

Moreover, although the proposed VoxelMorph-based cardiac motion extraction method can capture the frame-to-frame motion with sufficient accuracy, as shown in this work, our ongoing and future efforts are focused on further improving the algorithm by imposing diffeomorphic deformations \cite{dalca2019unsupervised}. This improvement will help maintain a high quality of the meshes and prevent mesh tangling and element degeneration, especially for the systolic phases. 

\vspace{-0.3cm}
\section{Conclusion}

In this work, we show that the proposed deep learning framework can be used to build LV myocardial geometric models. The proposed framework is not limited to any pathology and can be extended to LV and RV blood-pool geometry.  

\vspace{-0.3cm}
\section*{Acknowledgments}  

This work was supported by grants from the National Science Foundation (Award No. OAC 1808530, OAC 1808553 \& CCF 1717894) and the National Institutes of Health (Award No. R35GM128877).
\vspace{-0.4cm}

\bibliographystyle{splncs04}
%\bibliography{main}

\end{document}